\documentclass{article}

\PassOptionsToPackage{numbers, compress}{natbib}
     \usepackage[final]{neurips_2022}

\usepackage[utf8]{inputenc} % allow utf-8 input
\usepackage[T1]{fontenc}    % use 8-bit T1 fonts
\usepackage{hyperref}       % hyperlinks
\usepackage{url}            % simple URL typesetting
\usepackage{booktabs}       % professional-quality tables
\usepackage{amsfonts}       % blackboard math symbols
\usepackage{nicefrac}       % compact symbols for 1/2, etc.
\usepackage{microtype}      % microtypography
\usepackage{xcolor}  % colors
\usepackage{wrapfig}
\usepackage{graphicx}
\usepackage{algorithmic}
\usepackage{algorithm}
\usepackage{wrapfig}
\usepackage{lipsum}
\usepackage{amsmath}
\usepackage{hyperref}

\title{Normative Modeling using Multimodal Variational Autoencoders to Identify Abnormal Brain Structural Patterns in Alzheimer Disease}

\author{%
  Sayantan Kumar \\
  Department of Computer Science\\
  Washington University in St. Louis\\
  St. Louis, MO 63130 \\
  \texttt{sayantan.kumar@wustl.edu} \\
  \And
  Philip R. O. Payne \\
  Department of Computer Science\\
  Washington University in St. Louis\\
  St. Louis, MO 63130 \\
  \texttt{prpayne@wustl.edu} \\
  \And
  Aristeidis Sotiras \\
  Department Radiology\\
  Washington University in St. Louis\\
  St. Louis, MO 63130 \\
  \texttt{aristeidis.sotiras@wustl.edu} \\
}

\begin{document} 
\maketitle

\begin{abstract}

Normative modelling is an emerging method for understanding the underlying heterogeneity within brain disorders like Alzheimer Disease (AD) by quantifying how each patient deviates from the expected normative pattern that has been learned from a healthy control distribution. Since AD is a multifactorial disease with more than one biological pathways, multimodal magnetic resonance imaging (MRI) neuroimaging data can provide complementary information about the disease heterogeneity. However, existing deep learning based normative models on multimodal MRI  data use unimodal autoencoders with a single encoder and decoder that may fail to capture the relationship between brain measurements extracted from different MRI modalities. In this work, we propose multi-modal variational autoencoder (mmVAE) based normative modelling framework that can capture the joint distribution between different modalities to identify abnormal brain structural patterns in AD. Our multi-modal framework takes as input Freesurfer processed brain region volumes from T1-weighted (cortical and subcortical) and T2-weighed (hippocampal) scans of cognitively normal participants to learn the morphological characteristics of the healthy brain.  The estimated normative model is then applied on Alzheimer Disease (AD) patients to quantify the deviation in brain volumes and identify the abnormal brain structural patterns due to the effect of the different AD stages. Our experimental results show that modeling joint distribution between the multiple MRI modalities generates deviation maps that are more sensitive to disease staging within AD, have a better correlation with patient cognition and result in higher number of brain regions with statistically significant deviations compared to a unimodal baseline model with all modalities concatenated as a single input. 

\end{abstract}

\begin{figure}[htbp]
    \centering
    \includegraphics[width = 0.8\linewidth]{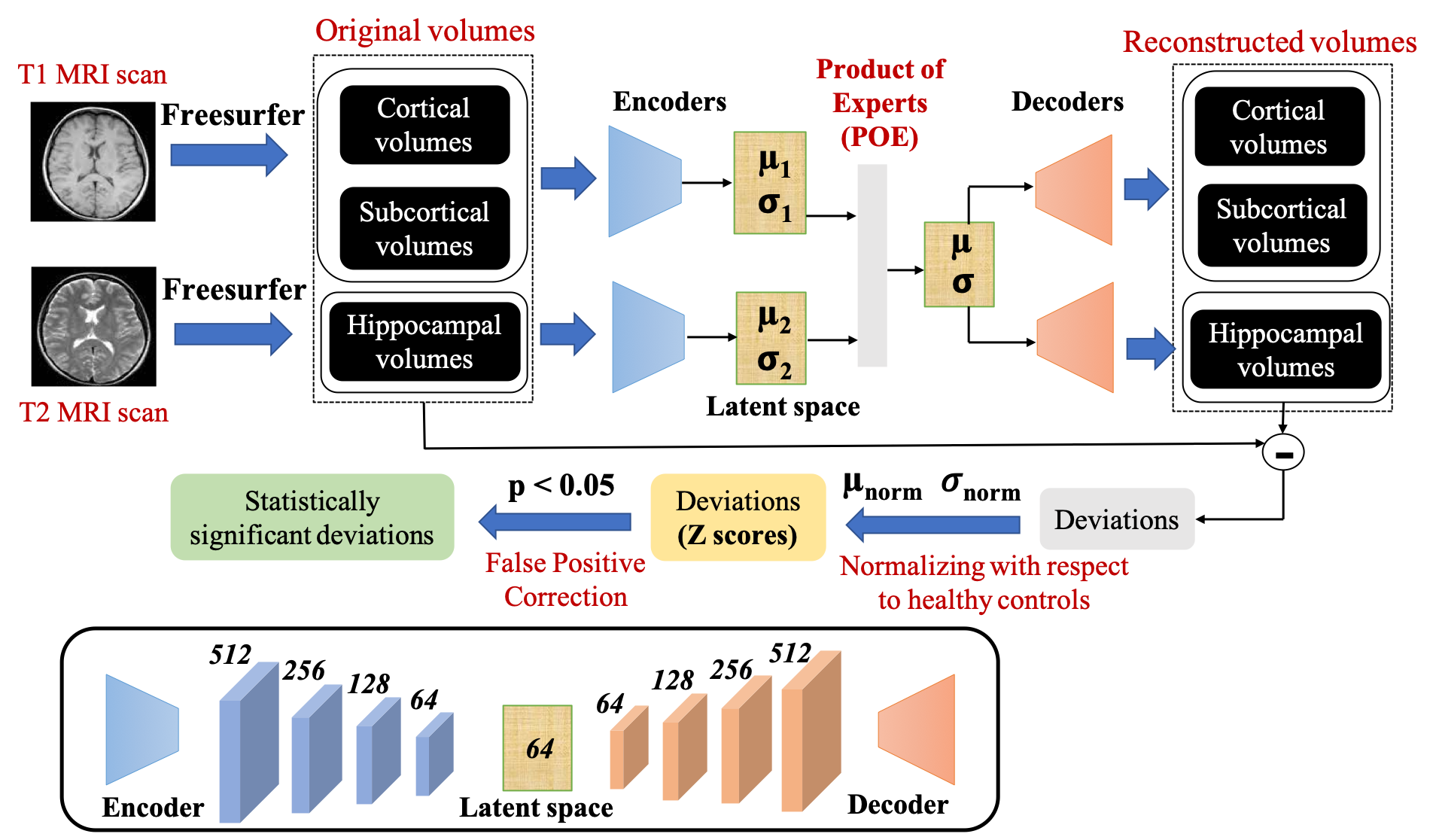}
    %\vspace{-10pt}
    \caption{Our proposed multimodal normative modeling framework (mmVAE). Cortical and subcortical brain volumes extracted from T1-weighted MRI scans (FreeSurfer) and hippocampal volumes extracted from T2-weighted MRI scans (Freesurfer) are used as input to two modality specific encoders. The encoder and decoder networks have 4 fully-connected layers of sizes {512, 256, 128, 64} and {64, 128, 256, 512}, respectively with a latent dimension of 64. The latent space parameters of the individual modalities are combined by the Product of Experts (POE) layer \cite{cao2014generalized,wu2018multimodal} to form the shared latent space, which feeds the modality-specific decoders for reconstructions. The deviations in brain volumes of disease patients are calculated, normalized with respect to the healthy controls to form Z-scores. The regions with statistically significant deviations are identified ($p_{FDR} < 0.05$) after correcting for false positives (False Discovery Rate). }
    %\vspace{-10pt}
    \label{fig:workflow}
\end{figure}

%\keywords{normative modeling, Alzheimer Disease, multimodal variational autoencoders}

\section{Introduction}
\label{sec:intro}  % \label{} allows reference to this section

Neurodegenerative disorders like Alzheimer Disease (AD) are highly heterogeneous, and the presence of different brain pathologies and variation in genetic background lead to significant variations in the clinical presentation and disease course. Traditional case-control analyses on AD assume that there is a single pattern that distinguishes the two contrasted groups and focus on 1st order statistics (group means) to estimate it, effectively ignoring the underlying disease heterogeneity \cite{dong2017heterogeneity, marquand2019conceptualizing}.Normative modeling is a technique for parsing heterogeneity in clinical cohorts, while allowing predictions at an individual subject level \cite{dong2017heterogeneity, marquand2019conceptualizing}. Assuming that neurodegenerative disorders like AD manifest as deviations from a normal pattern of functioning, the parameters of a normative model are learned such that they characterise healthy brains from a control population and provide a statistical measure of normality. Thus, applying the normative model to a disease cohort allows for quantification of the deviation of disease patients from the norm. \cite{marquand2016understanding, huizinga2018spatio,ziegler2014individualized,kia2019neural,bethlehem2020normative,wolfers2018mapping}. Regression models such as hierarchical linear models, support vector regression, and gaussian process regression (GPR) have traditionally been used as normative models (for an extensive list, see \cite{marquand2019conceptualizing}). However, it is necessary to train a regression model for each individual brain region which is not only computationally costly and also does not incorporate the interactions between brain regions. 

Given advances in deep learning technology and the growing availability of large scale datasets, there have been a number of deep learning-based normative models proposed in recent years that uses autoencoders to learn the morphological structure of healthy brains and subsequently used to analyze the brain pattern deviations of AD patients \cite{kumar2021normvae, pinaya2019using,pinaya2021using}. However, these studies train their models on a single modality of MRI data. Since AD is a multifactorial disease with more than one biological pathways, using multimodal neuroimaging data can provide complementary information about the biological mechanism of AD. The existing autoencoder-based normative models have unimodal structure with a single encoder and decoder that takes multimodal data concatenated as a single modality input. These unimodal frameworks do not take into account the correlations between different MRI modalities (e.g., T1-weighted and T2-weighted) which can potentially better capture the abnormal brain pattern deviations due to AD.\cite{ pinaya2019using, pinaya2021using}. 

\textbf{Contributions}: In this work, we aim at implementing a multi-modal variational autoencoder (mmVAE) based normative modeling framework that can capture the joint distribution between different MRI modalities to identify the abnormal brain structural patterns of AD patients. We use mmVAE as a normative model to learn the brain regional volumes of healthy subjects and subsequently quantify the deviation in brain volumes of AD patients. We hypothesize that deviations generated by our proposed multimodal framework are more sensitive to disease staging within AD, have a better correlation with patient cognition and estimate more regions with statistically significant deviations compared to a unimodal baseline model. To the best of our knowledge, ours is the first work on developing a multimodal normative modeling framework that can capture the joint distribution between T1-weighted and T2-weighted MRI modalities to identify abnormal brain pattern deviations due to the progressive stages of AD.

\section{Methods}

\subsection{Joint distribution between multiple modalities}

Our proposed mmVAE has separate modality-specific encoders and decoders for individual modalities. The main idea is to assume the joint distribution over the multiple modalities factorizes into a product of single-modality data-generating distributions when conditioned on the latent space. This assumption is used to derive the structure and factorization of the variational posterior. Without loss of generality, we assume that we have N modalities $x_1,..x_N$ which are conditionally independent given the common latent variable $z$. So, we assume we assume a generative model of the form $p_\theta(x_1,..x_N, z) = p(z) \prod_{i=1}^{N} p_\theta(x_i|z)$. The conditional independence assumptions in the generative model imply a relation among joint-modality posterior $p(z|x_1,..,x_N)$ and single-modality posterior $p(z|x_i)$ as shown below.

\begin{align*}
    p(z|x_1,..,x_N) &= \frac{p(x_1,..,x_N|z) * p(z)}{p(x_1,..,x_N)} = \frac{p(z)}{p(x_1,.., x_N)} * \prod_{i=1}^{N} p(x_i|z)\\
    &= \frac{p(z)}{p(x_1,..x_N)} * \prod_{i=1}^{N} \frac{p(z|x_i)p(x_i)}{p(z)} \\
    &= \frac{\prod_{i=1}^{N} p(z|x_i)}{\prod_{i=1}^{N-1} p(z)} * \frac{\prod_{i=1}^{N} p(x_i)}{p(x_1,..x_N)} \\
    &\propto \frac{\prod_{i=1}^{N} p(z|x_i)}{\prod_{i=1}^{N-1} p(z)}
\end{align*}

\subsubsection*{Approximating the joint posterior}

Here we see that the joint posterior is a product of individual posteriors, with an additional quotient by the prior. Alternatively, if we approximate $p(z|x_i) = q(z|x_i) \equiv \Tilde{q}(z|x_i)p(z)$ where $\Tilde{q}(z|x_i)$ the underlying inference network, we can avoid the quotient term $p(z)$. Now, we can approximate the joint posterior as shown below. In other words, we can use a product of experts (PoE), including a “prior expert” $p(z)$, as the approximating distribution for the joint-posterior

\begin{align*}
    p(z|x_1,..,x_N) \propto \frac{\prod_{i=1}^{N} p(z|x_i)}{\prod_{i=1}^{N-1} p(z)} \equiv \frac{\prod_{i=1}^{N} [\Tilde{q}(z|x_1) p(z)]}{\prod_{i=1}^{N-1} p(z)} = p(z) \prod_{i=1}^{N-1} \Tilde{q}(z|x_i)
\end{align*}

\subsubsection*{Products of Experts (PoE) approach}

Data from the individual modalities are fed into the corresponding encoders to form their respective latent space parameters (mean and variance). The product distribution required above are not in general solvable in closed form. However, if we approximate both $p(z)$ and $\Tilde{q}(z|x_i)$ as Gaussian, then we utilize the solution shown in \cite{cao2014generalized} that a product of Gaussian experts is itself Gaussian with mean $\mu = (\sum_{i} \mu_i*T_i)(\sum_{i}T_i)^{-1} $ and variance $\sigma = (\sum_{i}T_i)^{-1}$  where $\mu_i$ and $\sigma_i$ are parameters of the i-th Gaussian expert and $T_i = \sigma_i^{-1}$. 

% \begin{align}
% \mu = (\sum_{i} \mu_i*T_i)(\sum_{i}T_i)^{-1}  
% \sigma = (\sum_{i}T_i)^{-1}
% \end{align}

\subsection{Multimodal VAE (mmVAE) for Normative Modeling}

Although mmVAE can work for any $N$ number of modalities, we experimented with $N = 2$ in this work (Figure \ref{fig:workflow}). Brain volumes from the cortical and subcortical regions are extracted from T1-weighted Magnetic Resonance Imaging (MRI) scans by the FreeSurfer software (version 5.1) \cite{fischl2002whole}. Similarly, FreeSurfer was also used to extract the hippocampal region volumes from the T2-weighted MRI scans. The regional volumes from the T1-weighted and T2-weighted modalities were used as input to two modality-specific encoders to obtain their corresponding latent space parameters (mean and variance). The latent space parameters of the individual modalities were combined by the Product of Experts (PoE) approach as described above. As theoretically proved in the earlier section, the shared latent space formed by the PoE approach models the joint distribution between the two modalities. The combined latent values are then passed through modality-specific decoders to obtain the corresponding reconstructions. We also condition mmVAE on covariates like the age and sex of patients, represented as one-hot encoding vectors, to ensure that the deviations in regional brain volumes reflect only the disease pathology and not deviations due to effects of covariates (for details see Section 2.3)

\subsubsection*{Training on healthy controls}

mmVAE is first trained on the multimodal MRI (T1-weighted and T2-weighted MRI) of healthy controls (disease-free participants). In the training step, mmVAE learns to encode the healthy brain regional volumes of both modalities into a latent distribution and then uses the encoded representation to try to reconstruct the input (brain region volumes) as closely as possible to the original. Next mmVAE on patients with different stages of AD to assess the model sensitivity in estimating the extent to which they deviate from the normal trend. The main idea of the normative approach is that since mmVAE only learns how to reconstruct the brain region volumes of healthy controls, it will be less precise in reconstructing the brain volumes of AD patients. Hence, the difference between the actual and reconstructed data will be larger in AD patients compared to healthy controls.

\subsubsection*{Calculating the deviations}

For each disease patient, the deviations $d_{ij}$ with respect to the healthy controls are calculated as the absolute signed difference between the original and reconstructed brain region volumes. Assuming that the proposed model is not able to perfectly reconstruct the brain region volumes of the healthy subjects, the deviations are normalized with respect to the mean $\mu_{norm}$ and variance $\sigma_{norm}$ of deviations $d{ij}^{norm}$ of the healthy participants calculated from a separate held-out validation cohort. The final normalized deviation $deviation_{ij}$ (Z-scores) are calculated by normalizing the deviations of disease patients with respect to the mean $\mu_j$ and variance $\sigma_j$ of the deviations of healthy participants for each brain region $j$ (see equation below)
\begin{equation}
   deviation_{ij} = \frac{d_{ij} - \mu_{norm}(d_{ij}^{norm)}}{\sigma_{norm}(d_{ij}^{norm})} 
   \label{eq:deviation}
\end{equation}

We identified the brain regions of each patient whose deviations (Z-scores) are significantly different from those of the HC subjects ($p<0.05$). Since the normalized deviations $deviation_{ij}$ are estimated independently for each brain region for every patient, FDR (False Discovery Rate) correction was applied to control the Type 1 error rate (false positive correction) \cite{benjamini1995controlling}. Our proposed framework has been summarized in Figure 1.

\subsection{Experimental Design}

\subsubsection*{Data}

In our analysis, we used two datasets: the UK Biobank \cite{ sudlow2015uk} and the Alzheimer's Disease Neuroimaging Initiative (ADNI) \cite{mueller2005ways}. For training, we selected multimodal neuroimaging MRI data from 9633 healthy controls (HC) from UK Biobank after excluding all subjects with recent history of anxiety and depression, and subjects having previous hospitalization associated with diagnosis of mental and behavioral disorders, disease of the nervous system, cerebrovascular diseases, benign neoplasm of meninges, brain and other parts of the central nervous system. The estimated normative model on the UK Biobank was then applied on the multi-modal neuroimaging MRI data of 1136 patients from the ADNI dataset. To ensure data harmonization between UKB and ADNI participants, the normative model estimated on UKB was fine-tuned on ADNI by a transfer learning approach. The ADNI dataset included 269 cognitively normal (CN) controls and 862 patients: 106 Significant Memory Concern (SMC), 312 Early Mild Cognitive Impairment (EMCI), 263 Late Mild Cognitive Impairment (LMCI) and 181 AD patients. As patients progress from SMC to the AD stage, the severity of impairment increases. In the ADNI dataset, the MCI stages included participants with different levels of cognitive impairment with the EMCI groups showing milder impairment than the LMCI group.

\subsubsection*{Feature Preprocessing}

We used the FreeSurfer software (version 5.1) \cite{fischl2002whole} to estimate the brain regions’ volumes from the T1-weighted and T2-weighted MRI images respectively. This estimation was performed using the “recon-all” command (see \cite{fischl2002whole} for more information). From T1-weighted MRI scans, the cortical surface of each hemisphere was parcellated according to the Desikan–Killiany atlas \cite{desikan2006automated} and anatomical volumetric measures were obtained via a whole-brain segmentation procedure (Aseg atlas) \cite{fischl2002whole}. The final data included the cortical volume for each of the 64 cortical regions (32 per hemisphere), the volumes of 31 subcortical regions and volumes of 16 hippocampal regions.

As part of the feature pre-processing step, the brain region volumes of each subject were normalized by their Intracranial Volume (ICV). The healthy controls from UK Biobank data was split into 80\% for training and 20\% for validation, which was used for early stopping to prevent overfitting. The 269 CN participants in ADNI were split into 70\% for model training and validation, a 15\% held-out validation set for estimating the parameters of the normative population ($\mu_{norm}$ and $\sigma_{norm}$ in Equation \ref{eq:deviation}) and 15\% in the test for estimating the deviations. The volumes of each region were scaled between 0 and 1 using MinMax scaling. The mean and variance calculated in the training set were also used to scale the data in validation and test sets.

\subsubsection*{Conditioning on covariates (age and sex)}

We conditioned our proposed mmVAE on the age and sex of patients, to ensure that the deviations in regional brain volumes reflect only the disease pathology and not deviations due to effects of covariates. Both age and sex were transformed into one-hot encoding vectors. After this transformation, each subject has an age vector with 44 positions, where each position corresponds to a year within the range of 47–91 years. In this vector, all positions have value zero except the one that indicates the subject’s age which has a value equal to 1. The subject’s sex was represented in a one-hot encoded vector with two positions, one for male and one for female. Both the modality-specific decoders used these vectors together with the latent code to reconstruct the brain data.

\subsubsection*{Baselines and Implementation details}

We compared our proposed framework with a baseline VAE having a unimodal structure with a single encoder and decoder. The baseline framework takes the cortical, subcortical and hippocampal region volumes into a concatenated single-modality input. Similar to our proposed mmVAE, the baseline unimodal model is also conditioned on the age and sex of patients, represented as one-hot encoding vectors. Both the VAE models were trained using the Adam optimizer with model hyperparameters as follows: epochs = 500, learning rate = $10^{-5}$, batch size = 256 and latent dimension = 64. The encoder and decoder networks have 4 fully-connected layers of sizes {512, 256, 128, 64} and {64, 128, 256, 512}, respectively.

\subsection{Performance Evaluation}

We evaluate the deviations generated by both our proposed mmVAE and the baseline on the following three aspects : sensitivity towards (i) disease staging, (ii) correlation with cognition and (iii) identifying abnormal brain patterns.

\begin{itemize}
    \item \textbf{Sensitivity towards disease staging:} For each disease patient in the ADNI test set, we calculated the mean deviations (Z-scores) across all the 115 brain regions. The mean deviations of different categories of disease patients (SMC, EMCI, LMCI, AD) are analyzed to see if we have higher deviations across progressive stages (Figure \ref{fig:mean_deviation}).
    \item \textbf{Correlation with Cognition:} Pearson Correlation Coeffcient between the mean deviations (across all 115 regions) with 2 measures of cognition AD Assessment Scale (ADAS13) \cite{mohs1997development} and Rey Auditory Verbal Learning Test (RAVLT) \cite{matloubi2014effect} (Figure \ref{fig:correlation}). 
    \item \textbf{Abnormal brain deviation patterns:} For each of cortical, subcortical and hippocampal region, we calculated the frequency of significance which can be defined as the number of times in \% each region had statistically significant deviations from the healthy subjects ($p < 0.05$ after false positive correction) across all 862 patients in the test set. The regions with higher frequency of significance represent the abnormal brain structural patterns. (Figure \ref{fig:sig_freq}). 
\end{itemize}

\section{Results and Discussion}

\subsection*{Sensitivity of deviation maps towards disease staging}

\begin{figure}[htbp]
    \centering
    \includegraphics[width = \linewidth]{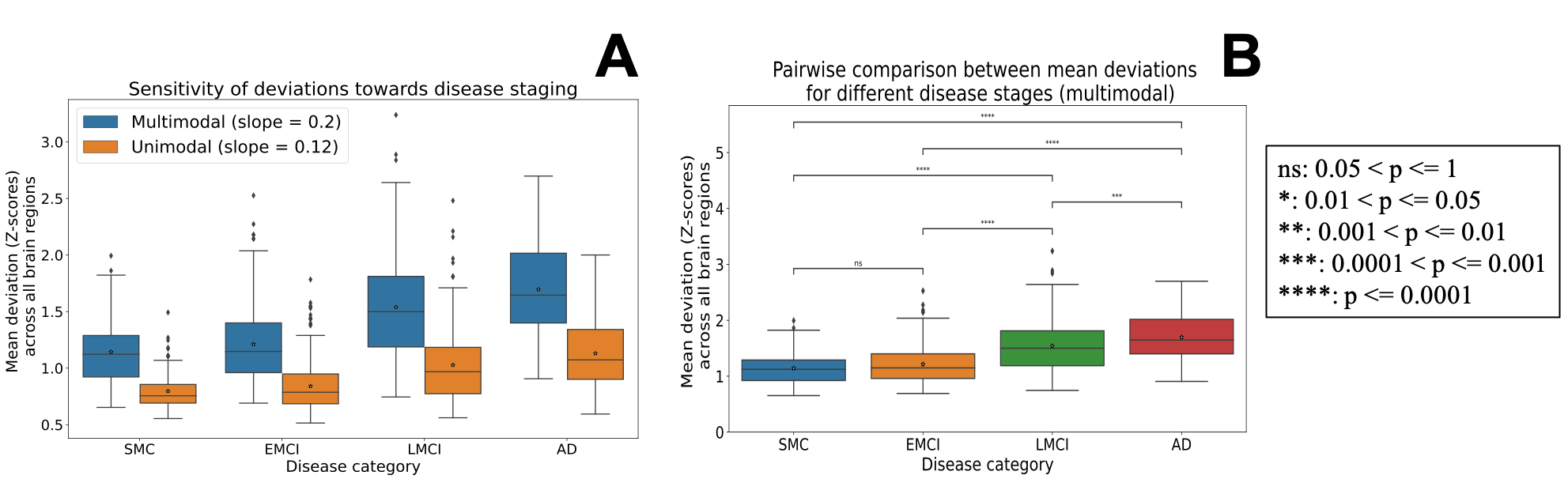}
    %\vspace{-10pt}
    \caption{\textbf{A}: Box plot showing the deviations (mean deviation across all $115$ brain regions) of each patient generated by our proposed and baseline models across all four AD disease categories. The slope values shown in the figure legend are obtained by fitting a linear model across the mean value of each category (shown by * within each box plot). \textbf{B}: Statistical annotations on the mean deviations generated by our proposed framework, showing the pairwise statistical comparison ($p_{FDR} < 0.05$) between the mean deviations in each disease category. Abbreviations: \textbf{SMC}: Significant Memory Concern, \textbf{EMCI}: Early Mild Cognitive Impairment, \textbf{LMCI}: Late Mild Cognitive Impairment, \textbf{AD}: Alzheimer Disease.}
    %\vspace{-10pt}
    \label{fig:mean_deviation}
\end{figure}

The mean deviations of each patient across all 115 brain regions reflect the measure of abnormality or neuroanatomical alteration in the brain due to the AD progression (SMC $\rightarrow$ EMCI $\rightarrow$ LMCI $\rightarrow$ AD) and should ideally include more regions with increasing severity of the disease stages. For both our proposed multimodal and baseline unimodal framework, the patients exhibited more abnormality (mean deviations) with increasing severity of their condition from SMC to AD (\textbf{Figure \ref{fig:mean_deviation}A}). For each model, the deviation slope across categories was calculated by fitting a linear model across the mean deviations in each category. Higher slope suggests that our proposed model is more sensitive to disease staging compared to the baseline. The pairwise differences in deviations between the disease categories were statistically significant ($p_{FDR}<0.05$) except for the SMC and EMCI pair (\textbf{Figure \ref{fig:mean_deviation}B}). Thus, the deviations generated by our proposed model can better capture the neuroanatomical alterations in the brain due to the progressive stages of AD. 

\begin{figure}[htbp]
    \centering
    \includegraphics[width = 0.7\linewidth]{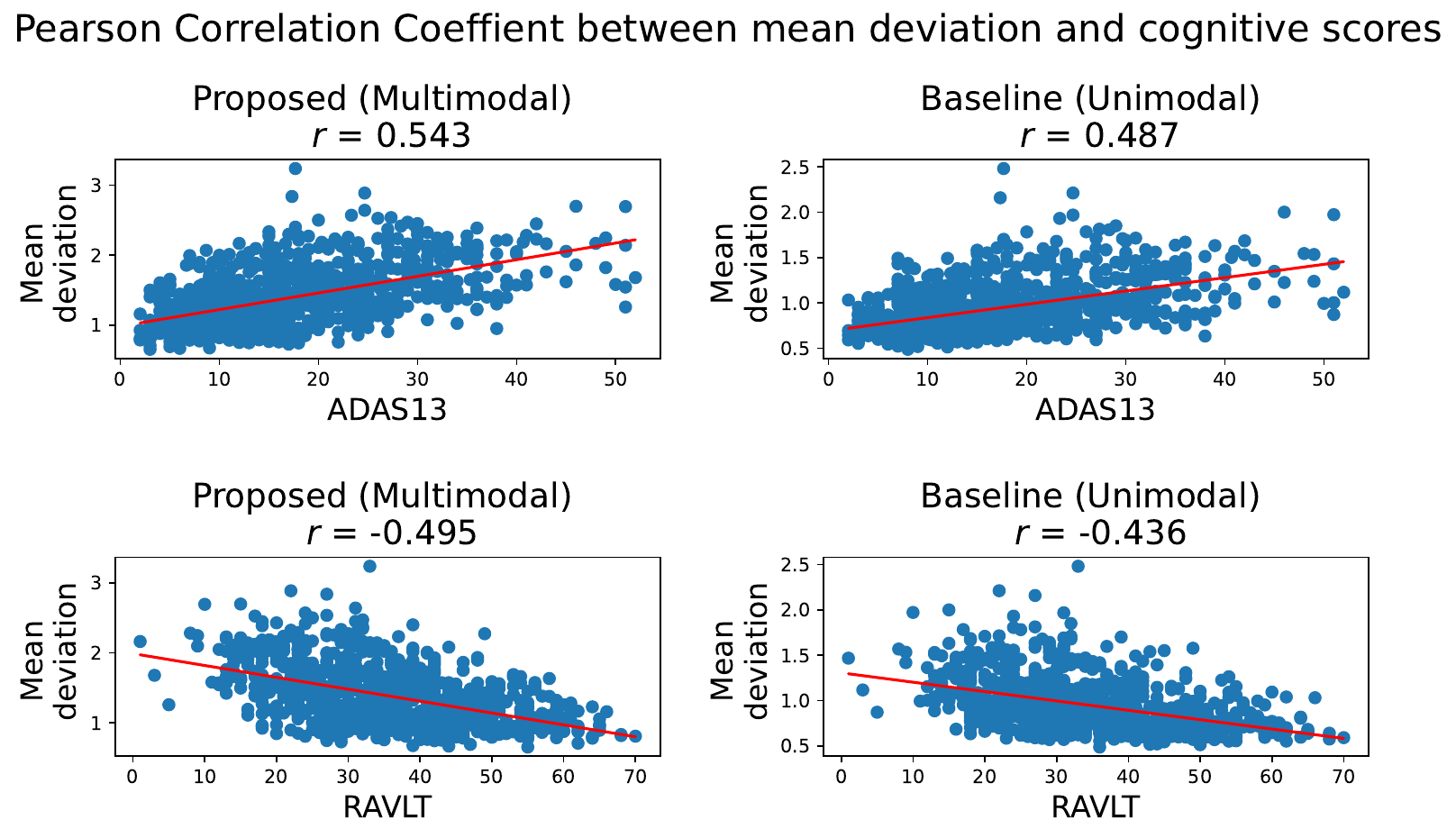}
    %\vspace{-10pt}
    \caption{Pearson Correlation between mean deviations (Z-scores) across all 115 brain regions and patient cognition represented by ADAS13 (top row) and RAVLT (bottom row). $r$ indicates the correlation coefficient value. Each point in the plot represents a patient and the dark red line denotes the linear regression fit of the points. Abbreviations: \textbf{ADAS13}: Alzheimer Disease Assessment Scale (ADAS13), \textbf{RAVLT}: Rey Auditory Verbal Learning Test.}
    %\vspace{-10pt}
    \label{fig:correlation}
\end{figure}

\subsection*{Correlation of deviation maps with patient cognition}

We analyzed the correlation between the patient-level deviations (mean deviatios across all 115 regions) and cognitive assessment scores, AD Assessment Scale (ADAS13) \cite{mohs1997development} and Rey Auditory Verbal Learning Test (RAVLT) \cite{matloubi2014effect}. ADAS13 test is a series of 13 cognitive tasks that can be used to assess the level of cognitive dysfunction in AD. RAVLT is a neuropsychological assessment to evaluate the nature and severity of memory impairment over time. High scores of ADAS13 and low scores of RAVLT indicate greater loss in memory and cognition. We calculated the Pearson Correlation coefficient between the patient-level deviations and ADAS13 and RAVLT (\textbf{Figure \ref{fig:correlation}}) with r indicating the correlation coefficient. Our proposed framework exhibited higher correlation with patient cognition, compared to the baseline ($p < 0.05$). 

\subsection*{Identifying abnormal brain deviations}

\begin{figure}[htbp]
    \centering
    \includegraphics[width = \linewidth]{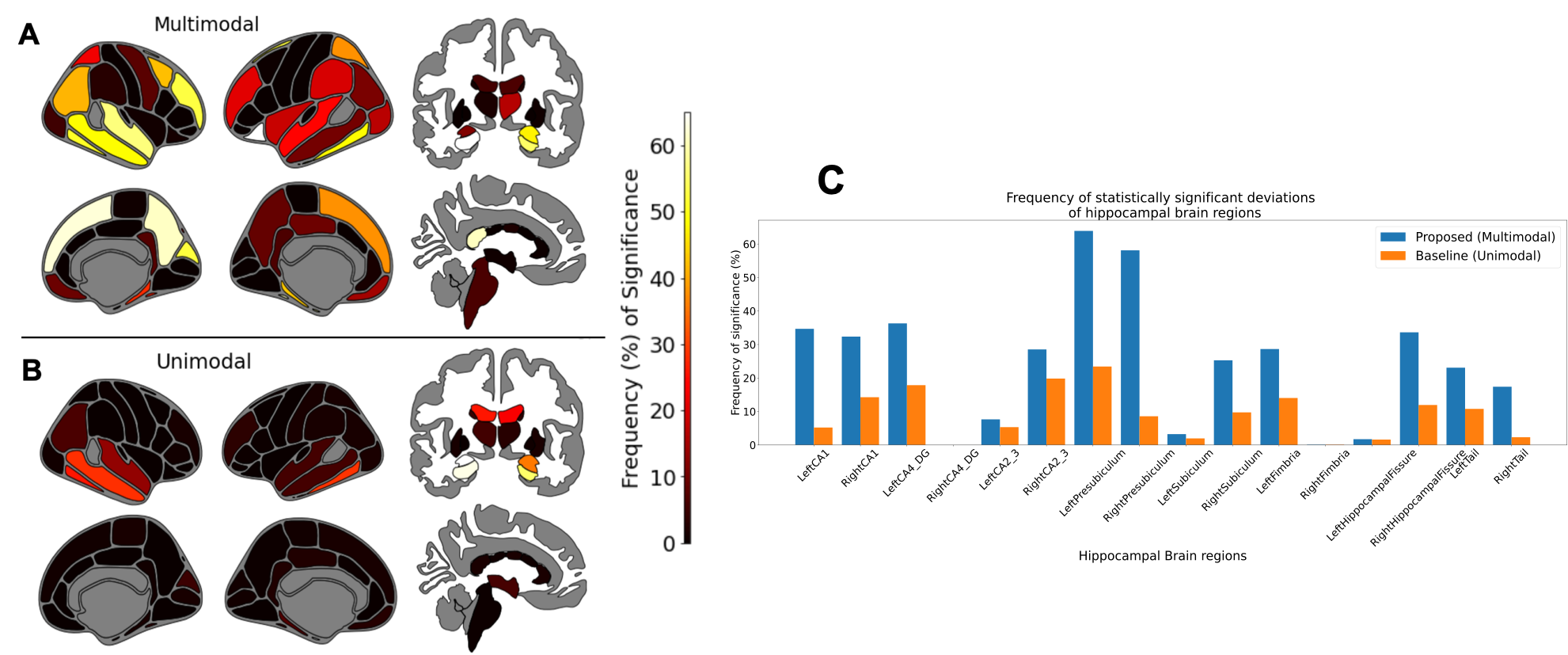}
    %\vspace{-10pt}
    \caption{Frequency of significance: Number of times ($\%$) each cortical and subcortical region (Figure \ref{fig:sig_freq}A, \ref{fig:sig_freq}B) and hippocampal region (Figure \ref{fig:sig_freq}C) exhibited statistically significant deviations ($p_{FDR} < 0.05$) across all 862 patients in the test set. The 1st 2 columns in Figure \ref{fig:sig_freq}A, \ref{fig:sig_freq}B represent the 4 views of the Desikan-Killany cortical atlas and the last column represent the coronal and saggital views of the Aseg subcortical atlas. The colormap shows the frequency of significance in \% with lighter colours representing higher frequency. The regions with higher frequency of significance across both the models (proposed multimodal and baseline unimodal) indicate the regions with abnormal brain deviation patterns due to AD.}
    %\vspace{-10pt}
    \label{fig:sig_freq}
\end{figure}

We identified the brain regions whose deviations in brain volumes compared to the healthy controls were statistically significant after false positive correction ($p_{FDR}<0.05$). For both the T1-weighted MRI and T2-weighted MRI modalities, we calculated how many times in \% each cortical, subcortical region (Figure \ref{fig:sig_freq}A, B) and hippocampal region (Figure \ref{fig:sig_freq}C) had statistically significant deviations across all 862 patients in the ADNI test set. The deviations generated by our proposed mmVAE has higher fraction of significance for all the regions compared to the unimodal baseline. 

The regions with higher frequency of significance across both the models (proposed and baseline) indicate the regions with abnormal brain structural patterns. In other words, we can get an idea which of the brain regions have more deviation in volumes due to the progressive stages of AD compared to the other regions. Some of the regions with showing deviations include (i) \textbf{cortical} : \textit{entorhinal-left, lateralorbitofrontal-left, fusiform-left, precuneus-right, superiorfrontal-right}, (ii) \textbf{subcortical} : \textit{hippocampus-left}, \textit{hippocampus-right}, \textit{amygdala-right}, \textit{Corpus-Callosum-Posterior} and (iii) \textbf{hippocampal} : \textit{presubiculum-left}, \textit{presubiculum-right}.

\section{Conclusion and Future Work}

In this work, we propose multi-modal variational autoencoder (mmVAE) based normative modelling framework that can capture the joint distribution between different modalities to identify abnormal brain structural patterns in AD. mmVAE takes as input Freesurfer processed brain region volumes from T1-weighted (cortical and subcortical) and T2-weighed (hippocampal) scans of cognitively normal participants to learn the morphological characteristics of the healthy brain.  The estimated normative model is then applied on AD patients to quantify the deviation in brain volumes and identify the abnormal brain structural patterns due to the effect of the different AD stages. Our experimental results show that modeling joint distribution between the multiple MRI modalities generates deviation maps that are more sensitive to disease staging within AD, have a better correlation with patient cognition and result in higher number of brain regions with statistically significant deviations compared to a unimodal baseline model with all modalities concatenated as a single input. 

As part of future work, we plan to perform further validations of our proposed model to estimate its generalizability on more neuroimaging datasets. Since the patient deviations are normalized with respect to those of the healthy controls, the mean deviation values for healthy controls in the ADNI test set should be ideally close to 0. However, it is probable to have false positives (high deviations of healthy controls in test set) due to mmVAE not being able to learn how to reconstruct data fully for ADNI CN during the re-training step. As a next step, we plan to validate this by adding some healthy controls in the test set and analyzing their deviations. 
The patient-level deviations generated by mmVAE can also be used for clustering patients into data-driven neuroanatomical AD subtypes. We believe that using the deviation maps will provide more interpretable clusters compared to just using brain region volumes alone. We also aim to test the discriminative power of the normative approach by using the deviation maps to classify AD patients vs healthy controls and compare the classification performance with traditional classifier taking only regional brain volumes as input. 

\subsection*{Data and Code availability}

UK Biobank data are available through a procedure described at \url{http://www.ukbiobank.ac.uk/using-the-resou rce/}. ADNI data are available through an access procedure described at \url{http://adni.loni.usc.edu/data-samples/ access-data/}. The full implementation code will be available soon in Github.

\subsection*{Acknowledgments}
This work was kindly supported by National Institutes of Health (NIH) grant number NIH R01-AG067103.

% References
\bibliographystyle{plain.bst}
\bibliography{references}

\end{document}